\documentclass[12pt,preprint]{aastex} 
\usepackage{graphicx}  
\newcommand{\mh}{H$_2$ }   
\newcommand{\mha} {H$_2$}

\newcommand{\s}{$\sim$}  
\newcommand{\bd}{\begin{displaymath}}  
\newcommand{\ed}{\end{displaymath}}  
\begin{document}  
\title{Formation and Evolution of Planetary Systems: Upper   
Limits to the Gas Mass in HD 105 \\}  
\author{D.Hollenbach \altaffilmark{1},  
U.Gorti \altaffilmark{2},  
M.Meyer \altaffilmark{3} 
J.S.Kim \altaffilmark{3}, 
 P.Morris \altaffilmark{4}, 
J.Najita \altaffilmark{5}, 
 I.Pascucci \altaffilmark{3}, 
 J.Carpenter \altaffilmark{6}, 
 J.Rodmann \altaffilmark{7}, 
 T.Brooke \altaffilmark{6}, 
L.Hillenbrand \altaffilmark{6}, 
 E.Mamajek \altaffilmark{8}, 
 D.Padgett \altaffilmark{4}, 
 D.Soderblom \altaffilmark{9}, 
S.Wolf \altaffilmark{7}, 
 J.Lunine\altaffilmark{10} } 
 
\altaffiltext{1}{NASA Ames Research Center, Moffett Field, CA 94035} 
\altaffiltext{2}{University of California, Berkeley, CA 94720} 
\altaffiltext{3}{Steward Observatory, The University of Arizona, Tucson, AZ 85721} 
\altaffiltext{4}{Spitzer Science Center, California Institute of Technology, Pasadena, CA 91125} 
\altaffiltext{5}{National Optical Astronomy Observatory, Tucson, AZ 85719} 
\altaffiltext{6}{California Institute of Technology, Pasadena, CA 91125} 
\altaffiltext{7}{Max-Planck-Institut f\"{u}r Astronomie, Heidelberg, Germany} 
\altaffiltext{8}{Harvard-Smithsonian Center for Astrophysics, Cambridge, MA 02138} 
\altaffiltext{9}{Space Telescope Science Institute, Baltimore, MD 21218} 
\altaffiltext{10}{Lunar Planetary Laboratory,  The University of Arizona, Tucson, AZ 85721}

\begin{abstract}

We report infrared spectroscopic observations of HD~105, a nearby ($\sim  
40$ pc) and relatively 
young ($\sim 30$ Myr) G0 star with excess  
infrared continuum emission, which has been modeled as arising from 
an optically thin circumstellar dust disk with an inner hole of size $\gtrsim 13$ AU. 
We have used the high spectral 
resolution mode of the Infrared Spectrometer (IRS) on  
the Spitzer Space Telescope to search for gas emission lines from the disk. 
The observations reported here provide  upper limits to the 
fluxes of H$_2$ S(0) 28$\mu$m,  
H$_2$ S(1) 17$\mu$m, H$_2$ S(2) 12 $\mu$m, [FeII] 26$\mu$m, [SiII] 35$\mu$m, 
and [SI] 25$\mu$m infrared emission lines. 
The H$_2$ line upper limits directly 
place constraints on the mass of warm molecular gas in the disk: 
$M({\rm H_2})< 4.6$, 3.8$\times 10^{-2}$, and $3.0\times 10^{-3}$ 
M$_J$ at $T= 50$, 100, and 200 K, respectively.  We also compare 
the line flux upper limits to predictions from detailed thermal/chemical 
models of various gas distributions in the disk.  These comparisons 
indicate that if the gas distribution has an inner hole with 
radius $r_{i,gas}$, the surface density at that inner radius is  
limited to values ranging from $\lesssim 3$ gm cm$^{-2}$ at $r_{i,gas}=0.5$ AU 
to  0.1 gm cm$^{-2}$ at $r_{i,gas}= 
5-20$ AU.   These values are considerably below the value for a minimum mass solar 
nebula, and suggest that  less than 1 M$_J$  
of gas (at any temperature) exists in the $1-40$ AU planet-forming region. 
  Therefore, it is unlikely that there is sufficient gas 
for gas giant planet formation to  occur in HD~105 at this time.  
\end{abstract} 
\section{Introduction}  
Observations of young stars indicate that circumstellar  disks of gas  
and dust are common   
in the earliest stages of evolution (e.g., Beckwith \& Sargent 1996,   
Hillenbrand et al. 1998, Haisch et al. 2001,  
Weinberger et al. 2004, Natta et al. 2004).   
Most of these studies rely on infrared, submillimeter, or millimeter  
wavelength emission as   
the signature of small ($\lesssim$ 1 mm) warm dust particles, which  trace   
only a small fraction of the circumstellar gas and solid particles.   
In their initial active  
accretion phase, circumstellar disks are composed   
of primordial molecular cloud material, possibly somewhat processed by  
the collapse through the accretion shock onto the disk,   
with gaseous  hydrogen  and helium constituting  
$99\%$ of the total mass and with $\sim 1\%$ of the disk mass in small   
dust particles. As infall from the cloud onto the  
star+disk system diminishes,   
the gas and dust components of the disk   
evolve, presumably losing most of the gas   
via accretion onto the central star or  
photoevaporation back to the interstellar medium   
(Hollenbach et al. 1994, Johnstone et al. 1998,   
Clarke et al. 2001, Adams et al. 2004).  Some of the small   
dust grains presumably coagulate,   
growing to eventually form planetesimals and   
rock/ice planets (e.g., Pollack et al. 1996,   
Weidenschilling 1977, 1997). During these processes,   
the circumstellar disk undergoes a transition  
from being optically thick and (mostly) gaseous to becoming optically thin and  
(mostly) dusty. There have been many theoretical and observational studies of   
disks in various stages of evolution pointing towards such a sequence   
(e.g., Weidenschilling 1997, Suttner \& Yorke 2001, Throop et al. 2001,   
Przygodda et al. 2003, Hogerheijde et al. 2003, Testi et al. 2003, van Boekel et al. 2003, 
Wolf et al. 2003, 
 Dullemond \& Dominik 2004, Kessler-Silacci et al.
2005).   
Such studies have, however,   
concentrated only on the more readily observed dust   
emission from the disk, and suggest the   
evolution of dust to larger particle sizes and hence lower opacity   
(Miyake and Nakagawa, 1993).   
  
Observations of gas and   
its evolution are in their infancy.   
Most studies to date have    
probed either at near infrared wavelengths (often CO vibrational transitions) the very hot   
inner surface regions less than about 1 AU from the star  
(Carr, Mathieu \& Najita 2001, Brittain et al. 2003, Najita et al. 2003, Blake \& Boogert 2004, 
Rettig et al. 2004, Thi et al. 2005),   
or at millimeter wavelengths (often CO rotational transitions)  extended  
gas--rich disks that  
are hundreds of AU in size   
(e.g., Skrutskie et al. 1991, Zuckerman et al. 1995, Dutrey  et al. 1998, Duvert et al. 2000,  
Pietu et al. 2003, Qi et al. 2003, , Dent et al. 2005, Greaves 2005).  
The millimeter observations are generally not sensitive to gas at   
$\lesssim$ 30 AU  
because the lines become beam--diluted and weak. In addition, the trace species used 
as millimeter wavelength probes of gas may freeze out onto grain surfaces at 
$\gtrsim$ 30 AU, making these observations somewhat unreliable for
estimating gas masses (e.g., van Dishoeck 2004). 

 
The presence  or absence of  
gas in the  planet-forming 1--30 AU regions is   
difficult to constrain from observations in the near infrared or at
millimeter wavelengths.  Temperatures in these  
regions may range from 300 --50 K, and warm gas emitting  
at these temperatures  is best probed in the   
mid--infrared lines of H$_2$ at 28.2 and 17.0 $\mu$m, and the mid--infrared  
fine structure lines of abundant ions and atoms, like [SI] 25.2 $\mu$m
(Gorti \& Hollenbach 2004, hereafter GH04).   
These lines are difficult or impossible to observe from the ground.    
Furthermore, the strong dust continuum of very young objects  
makes the line to continuum ratio quite low, and the background noise high,  
 unless the lines are spectroscopically resolved  
 (requiring resolving powers from 30,000--100,000).   
The Infrared Space Observatory (ISO) provided some tantalizing  
evidence for long--lived gas--rich disks by observing   
mid--infrared emission lines of H$_2$ toward debris disk candidates  
(Thi et al. 2001).   
However, ground--based observations (Richter et al.   
2002, Sheret et al. 2003, Sako et al. 2005), 
FUSE results (LeCavalier des Etangs et al. 2001), and Spitzer observations 
(Chen et al 2004) have called some of these  
results into question.  Recently, ground-based IR and   
space-based UV observations have   
detected rovibrational (e.g., Bary et al. 2003) and  
fluorescent (e.g., Herczeg et al. 2002) \mh emission towards young disks. 
The fluorescent \mh emission is found to occur from  
gas that is warm and probably located within a few AU  
of the star.  The narrow linewidths of the rovibrational IR emission  
suggests that the emission occurs at much larger distances from the star  
($>$10 AU), and arises either from non-thermal UV pumping or from  
the X-ray heating of a very small amount of \mh gas located in 
the low-density  
upper atmosphere.  The likely non-thermal excitation of the \mh seen  
both in the IR and UV makes it difficult to convert the detected line  
strengths into gas masses.

The nature of gas and dust evolution in circumstellar disks is an   
intriguing problem for various reasons. A determination of the gas dispersal   
timescales is important for understanding   
the formation of planetary systems, and  
specifically for discriminating between the two   
competing theories that exist for  
gas giant planet formation. Core accretion theories for forming gas giants  
 (e.g. Lissauer et al. 1993, Pollack et al.  
1996, Kornet et al. 2002, Hubickyj, Bodenheimer, \& Lissauer 2004) require long ($\gtrsim 
1$ Myr) gas disk lifetimes   
to facilitate the formation of  
rocky cores of a few Earth masses and their subsequent gas   
accretion to form planets.  
Detection of gas in older ($\sim 3 \times 10^6$ year) disks would help  
validate the efficacy of this scenario.   
The alternate theory of gravitational instability in   
disks (e.g., Boss 2003) allows  
the formation of gas giants quickly and the gas may dissipate rapidly.   
The presence of small amounts of gas in the  
``terrestrial'' zone, $0.1-5$ AU, in later evolutionary   
stages ($\sim$ 10 Myrs) influences   
the dynamics of the smaller planets and planetesimals and is decisive in the   
final outcome of the planetary system   
(Agnor \& Ward 2002; Kominami \& Ida 2002, 2004). Only a   
narrow range of gas masses ($\sim 0.01$ M$_J$)   
in this stage of terrestrial planet formation allow for   
the formation of an Earth-sized  
planet in an orbit as circular as the Earth. 
At   
even later stages ($> 10^7$ years)  
in disk evolution,   
small masses of gas ($\sim 0.01$ M$_J$) can affect dust   
dynamics via drag forces and  
hence significantly influence disk structure (Klahr \& Lin 2001,   
Takeuchi \& Artymowicz 2001, Takeuchi \& Lin 2002), weakening   
interpretations of gaps  
and ring signatures as being due to the presence of planets.   
  
This paper highlights the potential of the recently   
launched Spitzer Space Telescope  
in detecting small amounts of gas (or setting stringent limits on gas mass)  in   
the planet--forming regions of disks around nearby young stars.   
The Spitzer Space Telescope is able  
to observe the thermal emission from deep layers in the    
planet-forming zone, sampling significant mass in this 0.3--30 AU intermediate  
zone inaccessible to most other techniques. Gas in these regions is  
likely to   
be warm ($\sim 100$ K) and line emission at the mid-infrared ($10-37 \mu$m)   
wavelengths seen by the Spitzer Infrared Spectrometer (IRS) instrument  
can probe total gas masses as small as $\sim 0.01 $M$_J$ in the inner ($<$  
20 AU) regions around nearby ($\lesssim 30 $ pc) stars  
(GH04). Gas line diagnostics in the Spitzer band  
include the S(0), S(1) and  
S(2) pure rotational lines of \mha, atomic fine structure lines of [SI], [FeI],  
[SiII], [FeII] and molecular lines of OH and H$_2$O (GH04).   
  
The Formation and Evolution of Planetary Systems (FEPS) Spitzer Legacy   
Science Project  (Meyer et al 2005)
aims to study the evolution of gas in disks by targeting a carefully  
chosen sample of about 40 nearby solar-type stars for high   
resolution (R$\simeq$ 700) spectroscopic observations by the IRS (Houck et al.   
2004). The sample was selected mainly on the criteria that they span an age 
range of 3-100 Myr, and that they be nearby ($<140$ pc) so that the 
expected weak line fluxes may be detectable with the IRS. A secondary criterion 
was that they show some infrared excess emission indicative of the presence 
of  dusty disks.  To some extent, we favored sources with strong X ray luminosities, 
since the X rays heat the putative gas and intensify the resultant line luminosities. 
This will be the most comprehensive survey to date of gas  
in transition and post-accretion systems in order to characterize its dissipation and   
to place limits on the time available for giant planet formation.   
 
In this paper, we present high resolution IRS data and detailed modeling results for  
one of the first sources observed through the FEPS gas program, HD~105.    
HD~105 is a young ($\sim 30$ Myr) solar-type star 40 pc from the Sun   
 with an infrared excess indicative  of a   
dust disk (Meyer et al. 2004). The star has  completed its active  
accretion phase and the dust continuum is characteristic of an optically thin,  
debris disk. The youth and proximity of the star, the presence of dust (which  
may imply associated gas), the relatively low dust continuum luminosity that  
allows the detection of weak gas lines, and the presence of  X-ray flux  
from the central star   
(a source of gas heating) make HD~105 a promising candidate for the   
detection of mid-infrared gas lines.   
{\em Spitzer spectroscopic observations of HD~105,  
however, did not detect any gas emission lines, but instead set stringent  
upper limits on the line fluxes.}   
The line flux upper limits  are used to   
set  limits on the gas surface density and mass in HD~105.  
Assuming  
that the gas distribution has an inner hole of radius $r_{i,gas}$  
greater than about 0.5 AU, we find that there is  
very little gas in the disk beyond the inner hole, 
 suggesting the end of the main   
planet-building epoch for gas giants in these regions.   
  
Debris disks have been observed around early (e.g., 49 Cet, A1V) 
and late-type (e.g., AU Mic; MIV) stars and  are often  
characterized by their fractional infrared luminosities,   
$f=L_{IR}/L_{bol}$, which range from $10^{-5}-10^{-3}$. 
 HD~105 is thus a somewhat luminous debris disk, with a fractional  
infrared luminosity, $f=L_{IR}/L_{bol} \sim 4 \times 10^{-4}$.  
The disk may have evolved past its main planet formation stage, 
and is young enough to perhaps be a ``transition object'' with 
some residual, detectable gas. 
Though gas is readily detected in young disks (ages $\sim$ a few Myrs, 
and $f \gtrsim 0.01$, e.g. Dent et al. 2005), most debris disks do not appear 
to have detectable amounts of gas. Notable exceptions are the  
disks around 49 Cet($f \sim 10^{-3}$) and $\beta$ 
 Pictoris($f \sim 10^{-3}$). CO emission has been observed  
from the J$=3\rightarrow2$ (Dent et al. 2005) and the J$=2\rightarrow1$  
(Zuckerman, Forveille \& Kastner 1995) transitions from the disk around 
49 Cet, and ISO detected H$_2$ $28\mu$m emission as well (Thi et al. 2001).  
In $\beta$ Pic gas emission from metal atoms and ions orbiting in a  
disk has been reported, most recently by Brandeker et al (2004). 
They and Thebault \& Augereau (2005)  
derive gas masses in the disk of $\sim 0.1 - 0.4 $ M$_{\oplus}$. 
This amount of gas is insufficient to  brake the ions against radiation  
pressure; both groups estimate $\sim 40$ M$_{\oplus}$ 
is needed for braking.  Chen et al. (2004) obtain upper limits on
{\it warm} gas in $\beta$ Pic of about 11 M$_{\oplus}$, although
these Spitzer observations of the pure rotational lines of H$_2$
are not sensitive to the presence of cold ($< 50$ K) or atomic gas. 
The directly detected gas suggests  gas 
depletion in debris disks relative to the dust, in the few instances where  
gas has been observed. HD~105 is the first debris disk observed in our  
FEPS H$_2$ program aimed at studying gas dispersal in disks as they evolve. 
As we shall show below, depending on the gas temperature, Spitzer is 
able to detect gas masses $\gtrsim 1 - 10$ M$_\oplus$ for sources
as close as HD~105.

The outline of the paper is as follows.   
We describe the source (\S 2), and discuss  
the analysis of the Spitzer observations (\S 3). We then   
describe the gas and dust disk modeling (\S 4). Our results are  
discussed in \S 5 and the summary and conclusions presented in \S 6.

\section{Source Description}  
   
 HD~105 is a G0~V spectral type star (Houk 1978)  
 located at an Hipparcos distance of 40 $\pm$1 pc (Perryman et al 1997).   
 Based on its observed space motion and corroborating age  
diagnostics, Mamajek et al. (2004) conclude that it is a likely   
 member of the Tuc-Hor association.   
 The membership suggests an age of about 30\,Myr for HD~105.  
 There are various other indicators of the young age of this dwarf:  
 its Li I $\lambda$6707 equivalent width (e.g. Cutispoto et al. 2002)  
 is similar to the $\sim$50\,Myr-old members of the IC~2602 and 2391  
clusters (Randich et al. 2001),   
 and it is stronger than that of the ~120\,Myr-old Pleiades stars  
(Soderblom et al. 1990);  
 HD~105 has an active chromosphere as revealed by the detection of   
 Ca II H and K emission (Henry et al. 1996) and its   
 X-ray luminosity ($L_{\rm{X}}=2.4\times 10^{29}$\,ergs\,s$^{-1}$) 
suggests youth (Metanomski et al. 1998, Wichmann et al. 2003).   
 Based on all these age indications,  we assigned an age of 30$\pm$10\,Myr  
 for HD~105 (Meyer et al. 2004).  
 
 \begin{figure}  
\includegraphics[scale=0.5, angle=270]{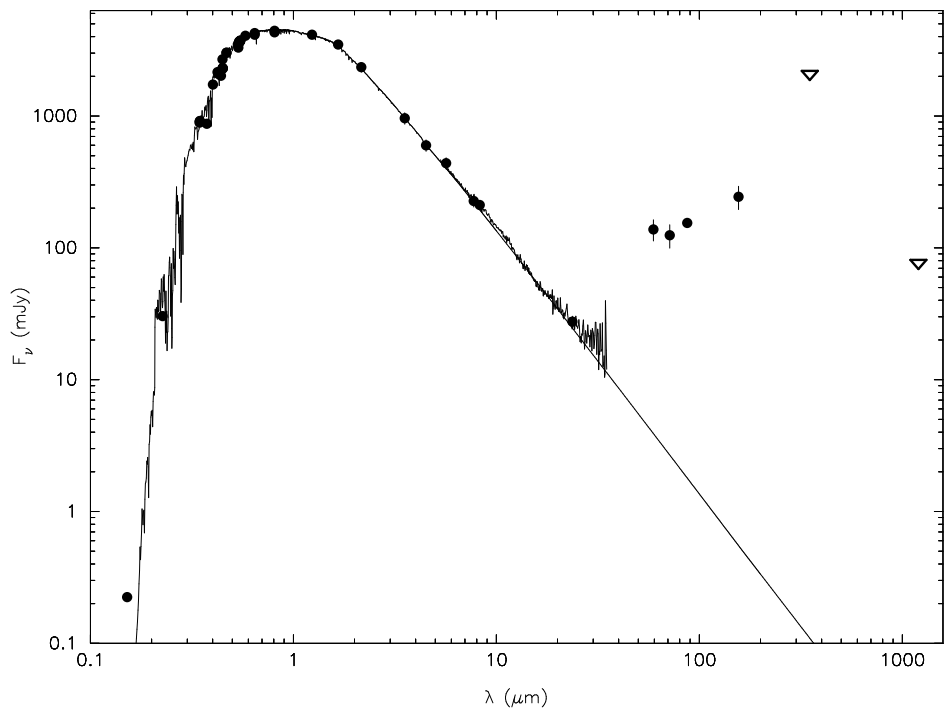}  
\caption{Observations and Kurucz model of HD~105. The solid curve through the observed 
fluxes is the best fit Kurucz model with $T_{eff}=6063$ K, $A_V=0.0$, and log~$g$ = 4.7. 
Also shown are the ISO data at 60 and 90 $\mu$m (Silverstone 2000),  
our own Spitzer MIPS and low resolution IRS data 
(Meyer et al. 2004), GALEX data taken from its archive,  
 a 350 $\mu$m upper limit provided by B. Mazin, and a 1 mm upper limit from 
Carpenter et al. (2005). }  
\label{SED}  
\end{figure}    
 Early ISO observations showed that HD~105 displays infrared excess emission at 60 and   
90 $\mu$m (Silverstone 2000). Our recent  
 Spitzer measurements detect no excess emission at 24$\mu$m,  
confirm the excess at 70 $\mu$m, and   
 detect dust emission even at 160 $\mu$m (Meyer et al. 2004, hereafter M04).   
B. Mazin (private 
communication) reports a 3$\sigma$ upper limit to the 350 $\mu$m continuum flux 
of 2.1 Jy.  
 We have searched the GALEX database at the position of HD~105 and  
found a source offset from the HD~105 position by about 3.2'' in the NUV, but  
only 0.09'' in the FUV, whereas the  
typical positional uncertainty is about 1''.  However, there is no other GALEX source within  
30'' of HD~105. Figure 1 shows 
the Kurucz model SED and the observed SED, including the infrared, submillimeter, and GALEX data.
As expected for a young star, the source has FUV excess.   
Meyer et al. (2004) presented early models of the dust distribution implied by  
the far infrared excesses.  The lack of detectable excess emission at wavelengths shorter than   
35 $\mu$m implies very little circumstellar dust in the inner ($\lesssim 15$ AU)   
disk.   We discuss these early models and other possible model  
fits in \S 4.

\section{Observations and Analysis} 
\subsection{Observations and Data Reduction} 
 
We obtained 9.9 - 37.2$\mu$m\ spectra of HD~105 on 
December 14, 2003 UT using the Infrared Spectrograph (IRS, Houck et  
al. 2004) 
onboard the Spitzer Space Telescope (Werner et al. 2004).   
The spectra we present here were obtained with the two echelle modules 
($R = 700$)  
covering 9.9 - 19.5$\mu$m\ with the Short High (SH) detector array,  
and 18.9 - 37.2$\mu$m\ 
with the Long High (LH) array.  We also acquired low resolution  
observations ($R=80$), which were used  
to account for background flux levels (discussed  
below). 
The spectra were acquired using the standard Staring 
Mode AOT, in which the telescope is nodded to place the star  
at two positions along each slit,  
resulting in 
four separate pointings (two for SH, and two for LH).  For the SH  
observations, we used integration times of 31.5 seconds for each  
Data Collection Event (DCE), and cycled eight times  
for a total on-source 
exposure time of 504 seconds.  For LH we used the 14.7 second  
integration time, also 
cycled eight times, for a total exposure time of 235 seconds.  High  
accuracy blue peak up ($13.3 - 18.7\mu$m) was performed on the star  
to ensure a pointing accuracy within  
0$''$.4.  The brightness profile of HD~105 in the peak up  
image agrees with that expected from the point spread function. 
 
Individual DCEs were processed through the SSC pipeline (version  
S10.5.0) resulting in Basic Calibrated Data (BCD) products to which  
basic detector calibrations, dark current 
subtraction, cosmic ray detection, integration ramp fitting, and  
reduction to individual, 
two-dimensional slope images in units of electrons/second/pixel 
have been applied.  
Dark current measurements taken with the LH module within 12 hours of  
the HD~105 observations 
were used to mitigate the effects of outlier pixels whose dark  
currents vary on 
timescales of one to several days, rather than using standard 
dark current subtraction  which employs darks taken over a wider  
separation in time. 
 
The number of LH outlier pixels (high dark current pixels  
not representative of expected gaussian noise)  
is less than 6\% of the 128$\times$128 pixels on the array.  However,  
they may dominate the noise within each echelle order if  
untreated.   
Corrections were applied to pixels flagged as anomalous in 
the pipeline, due to cosmic-ray saturation early in  
the integration, or having been preflagged as unresponsive.  
For these pixels we 
applied cubic spline interpolation technique, relying on 1.5  
resolution elements on either side of 
the dead pixels in the dispersion direction.  This signal  
reconstruction method has been successfully 
verified with observations of spectral standards exhibiting resolved  
and unresolved emission lines, particularly for the LH array.  
 
Individual 1-dimensional spectra were extracted from two dimensional  
images flat--fielded with observations of the zodiacal background.  
We used the offline SSC software to extract these spectra using the  
full width of each echelle order.  Spectral response 
and flux calibration corrections were applied to these 1-dimensional  
spectra.  The response functions and flux calibration 
are based on similarly processed and spatially flatfielded 
observations of photometric standards HR~6688 (K2III), HR~6705  
(K1.5III), HR~7310 (G9III), and HR~2194 (A0V), 
and the MARCS-code stellar atmosphere models tailored for these stars  
(Decin et al. 2004).  The response 
correction of the short wavelength end of the SH data ($9.9 -  
12.0\mu$m) of HD105 relied solely on HR~2194, 
in order to minimize any possible discrepancies in the strength of the  
SiO fundamental band in the K giants.   
 
The 16 individual spectra for each module were next median combined 
on a spectral order basis. Each order was  
trimmed, removing $\sim$5\% of the data at the blue ends where the  
throughput response of the arrays is lowest.   
The signal-to-noise in the continuum is lower than the signal-to-noise  
that would be inferred 
from repeatability among the 16 spectra. This suggests that 
the signal-to-noise is limited by residual errors due to 
fringing, uncertainty in the spectra response function and  
anomalous pixels among others. 
 The errors in absolute fluxes, which here include both  
the star and zodiacal background, are thought to be $\sim$ 20 \%.  
 
Although we did not obtain off-source exposures with SH or LH for sky  
subtraction, approximate corrections were applied, based on the background  
estimator available in the SPOT planning tool.   
These fluxes are provided in MJy/sr, which we then converted to Jy 
using the solid angles subtended by each of SH and LH apertures (1.25  
$\times$ $10^{-9}$ sr and 5.82 $\times$ $10^{-9}$ sr, respectively). 
Since the SPOT background estimate is derived from measurements made  
with low spatial resolution, we also compared the extracted sky spectrum  
obtained with the IRS near HD 105 at low spectral resolution.  
We found agreement to within 20\% of the fluxes provided by  
the background estimator between 15 and 40 $\mu$m\,  
when scaled by the appropriate solid angles.  
 
\begin{figure} 
\includegraphics[scale=0.7,angle=90]{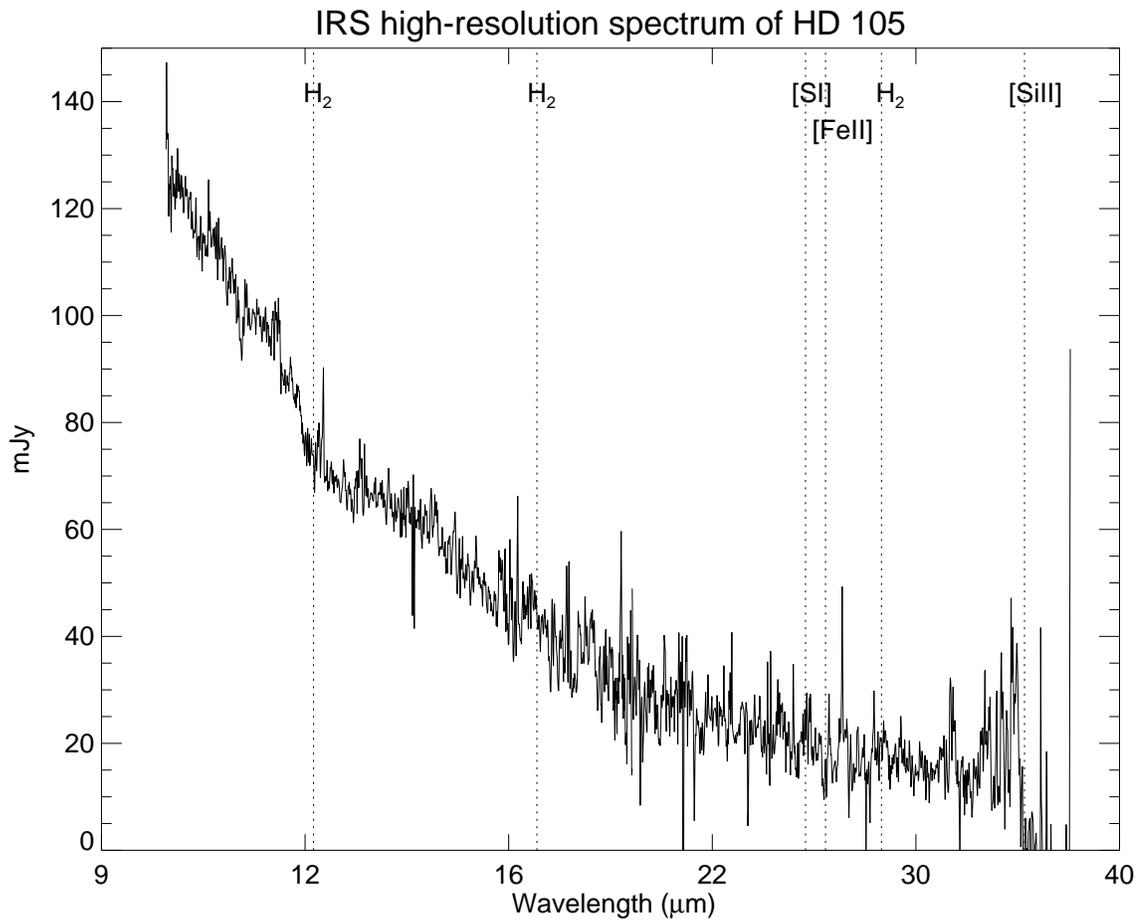} 
\caption{IRS spectrum  of the 10$\mu$m to 38$\mu$m wavelength 
covered by SH and LH, the high resolution mode.} 
\label{spectrum} 
\end{figure}

\begin{figure} 
\includegraphics[scale=0.7,angle=90]{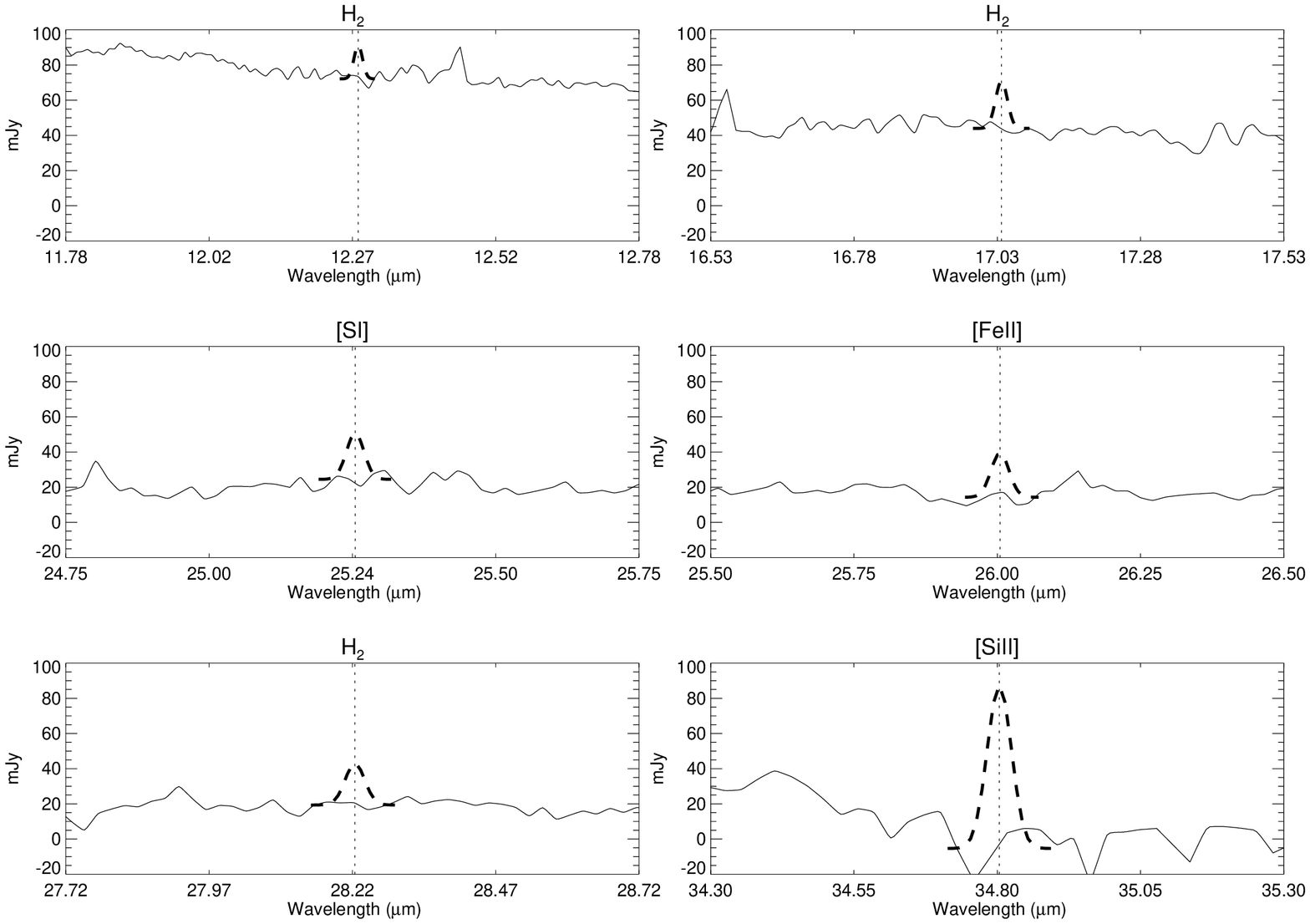} 
\caption{Expanded view of the wavelength regions around 
the expected strong lines. 
 Also shown in dashed lines are the hypothetical 5~$\sigma$  
line fluxes.} 
\label{spectrum2} 
\end{figure}

Figure~\ref{spectrum} shows the high  
resolution spectrum of HD~105.   
Figure~\ref{spectrum2} highlights the regions around the 
strongest emission lines expected from our models 
(Gorti \& Hollenbach~2004). 
The S/N is approximately 12 in  
the 16 -- 18$\mu$m\ range, where the noise 
is computed as the 1-$\sigma$ RMS spread in the ratio between the 
sky-subtracted IRS spectrum and the adopted Kurucz photospheric  
model over this essentially featureless 
region of the continuum  
(excluding the 16.9 -- 17.1 $\mu$m range where we  
searched for evidence of the H$_2$ S(1) line).   
Similarly, we estimate a S/N ratio 
of approximately 10 in the 26.5 -- 28.5um range.   
The S/N ratio becomes very low at $\lambda > 35 \mu$m, where the LH  
throughput is at its lowest.

\subsection{Analysis} 
  
To estimate upper limits of unresolved lines, we fit the 
spectrum using a Levenberg-Marquardt (LM) algorithm assuming a Gaussian 
for the line profile and a second order polynomial for the continuum.  
The central wavelength of the line was fixed in the fit, as was the 
width of the Gaussian to match the instrumental line profile.  
To fit the baseline locally, we used 
a wavelength range of $\pm$ 0.5$\mu$m centered  
on each feature (see Table~\ref{line-fluxes}).  
Five sigma upper limits to the line flux for each non--detection were derived  
taking the local RMS dispersion in the averaged   continuum over two  
pixels per resolution element.  We show  examples of  hypothetical 
5$\sigma$ lines in Figure 3.  
 Several 
anomalous pixels are evident in the spectra (e.g. at 12.45 and 16.56\,$\mu$m) 
that were not flagged in the data reduction discussed in \S 3.1. 
Since these do not correspond to any expected emission lines, 
this suggests that residual systematic effects may be present in the spectra  
shown. 
Table~\ref{line-fluxes}  gives the upper limits to the line fluxes of 
H$_2$, [SI], [FeII], and [SiII] lines to 
estimate upper limits of gas mass in this system. 
 
\begin{table}[bt] 
\caption{Line flux upper limits obtained by Spitzer IRS High Resolution 
observations of HD105} 
\begin{tabular}{lccc} 
\hline 
Gas Species & Wavelength   & Line flux upper limits \\ 
 &  ($\mu$m) &   (W/cm$^2$)\\ 
\hline 
\mh S(2) & 12.28 &  7.1 $\times 10^{-22} $\\ 
\mh S(1) & 17.035 &  7.2 $\times 10^{-22} $\\ 
SI  & 25.249 &  4.7 $\times 10^{-22}$\\ 
FeII  & 26.000 &  4.4 $\times 10^{-22}$\\ 
\mh S(0) & 28.221 &  3.8 $\times 10^{-22}$\\ 
SiII  & 34.800 &  1.2 $\times 10^{-21}$\\ 
\hline 
\end{tabular} 
\label{line-fluxes} 
\end{table}

\section{Modeling the Gas and Dust in the Disk Around HD105}  
\subsection{Simple Estimate of the Upper Limit to Warm H$_2$ Mass}  
 
Before presenting detailed models of gas/dust disks around HD~105,  
which can provide estimates of upper limits to the {\it total} 
gas mass in atomic and molecular form and at a (model--determined)  
range of temperatures, we calculate here a simple upper limit to  
the warm H$_2$ mass assumed to be at a constant temperature $T$  
in the HD~105 disk.  We use the upper limits  
on the H$_2$ S(0) 28$\mu$m and S(1) 17$\mu$m line fluxes to directly  
set limits on the mass of warm H$_2$ at temperature $T$ in HD~105.  
  
The H$_2$ S(0) and S(1) transitions have critical densities much  
lower than the gas densities in disks with gas masses $\gtrsim 3\times  
10^{-5}$  
M$_J$, the extreme lower limit for H$_2$ line detection by Spitzer  
of HD~105 or other sources at 40 pc (see below).   
Therefore, for any detectable disk or even for disks with H$_2$ masses 
somewhat below detectability, the lower rotational levels (i.e., J=0-3) 
of  H$_2$ are in LTE.   
  
With LTE and optically thin assumptions  
we calculate the mass $M(H_2)$ of molecular  
gas at temperature  
$T$ and at the distance (40 pc) of HD~105 that will produce a line flux  
$F$ for both the S(0) and S(1) transitions.  
  
\begin{equation}  
M(H_2) \simeq 2.8 \times 10^{-5} \left({F_{S(0)}   
\over {10^{-22}\ {\rm Watts\ cm^{-2}}}}  
\right)\left( 1 + {T \over {85\ {\rm K}}}\right)e^{510\ {\rm K}/T}\ M_J  
\end{equation}  
  
\begin{equation}  
M(H_2) \simeq 7.4 \times 10^{-7} \left({F_{S(1)}   
\over {10^{-22}\ {\rm Watts\ cm^{-2}}}}  
\right)\left( 1 + {T \over {85\ {\rm K}}}\right)e^{1020\ {\rm K}/T}\ M_J  
\end{equation} 
Utilizing these equations and the upper limits for the line fluxes given  
in Table~\ref{line-fluxes},   
we find upper limits on the H$_2$ gas mass of 4.6 M$_J$ at  
50 K, 3.8$\times 10^{-2}$ M$_J$ at 100 K, and 3.0$\times 10^{-3}$ M$_J$  
at 200 K.  The S(0) line flux sets the limit at 50 and 100 K, whereas  
the S(1) line flux sets the limit at 200 K. The S(2) line does not 
set useful limits at these low temperatures, where it is very weak. 
Note that assuming 
long Spitzer integrations that set H$_2$ S(1) flux  
limits of order $\sim 10^{-22}$ 
W cm$^{-2}$, the minimum H$_2$ mass detectable at 40 pc via  
the S(1) line is $3\times 10^{-5}$ M$_J$,  
achieved when the gas is $T \sim 1020$ K 
(see Eq. 2).  Equations (1 and 2) simply relate the flux in
an H$_2$ line to the mass of gas at a particular temperature.
In the following section, we apply much more sophisticated models
which actually calculate the temperature distribution in
a disk, and relate the fluxes to the total mass of gas
distributed in the disk.
 
\subsection{Thermal/Chemical Modeling of Gas and Dust}  
\subsubsection{Dust modeling}  
The Spectral Energy Distribution (SED) obtained from existing  
data in the literature and new Spitzer observations (see Figure 1)  
has been modeled  
using the dust disk models of Wolf \& Hillenbrand (2003).   
The  initial dust disk model for HD105 was presented  
in M04  
and we only give a brief description of these previous  
 results here. The   
observed infrared continuum excess cannot be uniquely fit by any one    
particular dust model but by a range of dust parameters.   
 M04 assumed for simplicity that  
the dust is composed of astronomical silicates with a surface density   
distribution $\Sigma(r) \propto r^0$. The model fits  were relatively insensitive  
to the exponent in the radial density distribution and the outer disk radius  
$r_{o,dust}$, since much of the dust emission at wavelengths shorter than the 
peak emission arises from dust 
near the inner radius, $r_{i,dust}$. A grain size distribution of $n(a) \propto a^{-s}$ with 
$s=3.5$ was chosen,   
representative of  regions with grain shattering, and providing  
somewhat better fits to the data than a single grain size.   
With such  
a distribution, most of the dust mass is in the largest particles, while the  
source of the infrared emission derives from grains with  the  
minimum size $a_{min}$ (which holds most of the dust area). We define  
``dust'' to be particles with sizes less than $a_{max}=1$ mm, and this definition  
sets the dust mass for a model fit. The three main parameters which determine the  
fit to the observed SED are then the inner radius $r_{i,dust}$, the minimum  
grain size $a_{min}$, and the dust mass $M_{dust}$.  
Roughly speaking, for a given $r_{i,dust}$ and $a_{min}$, the dust mass 
sets the total solid angle subtended by dust grains, and therefore sets the 
ratio $L_{IR}/L_{bol}$ of the dust IR luminosity to the stellar bolometric 
luminosity (an observed quantity which then determines the dust mass). 
M04 found acceptable fits with $r_{i,dust}= 32$ AU and $a_{min}= 8\ \mu$m, 
and with  $r_{i,dust} = 45$ AU and $a_{min}= 5\ \mu$m.   
These best 
fit $r_{i,dust}$ and $a_{min}$ corresponded to dust masses of 9$\times 10^{-8}$ 
and $4\times 10^{-7}$ M$_\odot$.  
  
Since the results presented in M04, we have performed a more extensive parameter study of the best  
$\chi^2$ fit, and have found that the minimum in $\chi^2$ corresponds  
to even larger $a_{min}=21$ $\mu$m  and lower $r_{i,dust}=19$ AU than explored in M04, and to  
somewhat lower dust mass \s $7.5 \times 10^{-8}$ M$_{\odot}$.   
We found that 
quite good fits could be obtained for 13 AU $< r_{i,dust} \lesssim 45$ AU. 
 The 
inner dust radius cannot be smaller than 13 AU because the dust becomes too 
hot, and produces excess emission at $\lambda \lesssim 35$ $\mu$m. 
We do not emphasize this SED fitting procedure here because we have found, and show below, that 
for all acceptable dust models, there is too little dust surface area to affect 
the gas chemistry, heating, or cooling.  Therefore, the gas spectrum is independent 
of the dust in the case of HD~105, and mainly depends on $r_{i,gas}$ and the gas 
surface density $\Sigma _0$ at $r_{i,gas}$, as we discuss below.

\subsubsection{Gas/dust disk models}  
We apply the thermal/chemical  disk models of Gorti \&  
Hollenbach (2004) to study  HD~105. In these models, a central 
star and the interstellar radiation field illuminate a gas and 
(optically thin) dust disk extending from $r_i$ to $r_o$.  The gas  
and dust are heated by the radiation field (the gas is particularly 
sensitive to the UV and X-ray radiation), and  
the gas and dust temperatures are calculated in separate 
thermal balance equations. The   
vertical density structure and chemistry is self-consistently   
computed by imposing thermal balance, steady-state chemistry  
 and pressure equilibrium.  The assumption in the dusty regions  
is that the gas and small dust particles are well mixed, so that  
their density ratio does not vary vertically (no settling of small  
dust particles).  However, as we shall show, there is so little  
dust in HD~105 that the gas emission lines do not depend on  
the dust vertical (or radial) distribution.   
We consider various gas heating sources such as gas-dust collisions,   
X-rays, stellar and interstellar FUV radiation, and exothermic chemical   
reactions. The  cooling is mainly by molecular 
rotational and atomic and ionic fine structure emission. 
In summary, the inputs to the model include the stellar parameters, 
the interstellar field, $r_i$ and $r_o$,  the surface density 
distribution of the gas, the dust size distribution, and the dust-to-gas 
mass ratio (which often is held fixed). Our main input variable 
generally is the gas surface density distribution.  The output 
is the vertical density, chemistry, 
 and temperature distribution as a function of $r$, 
and the resultant line intensities of various ionic,  
atomic and molecular species.  In turn, these line intensities 
can be compared with line observations to constrain the physical parameters 
in the disk, such as the gas surface density distribution (or in this 
case provide upper limits of gas surface densities). 
 
 
       We first consider a standard case, our best fit dust distribution 
discussed in \S4.2.1, with $r_{i,gas}=r_{i,dust}=19$ AU, i.e.,  
{\em gas and dust  co-exist  
spatially}.  However, we have found in this standard 
case, and in other SED-fitting cases we have tested  
with co-existing dust where $r_{i,dust} = 13-45$ 
AU, that the dust has no effect on the gas properties.  There is too 
little surface area to appreciably affect the cooling (through gas-dust 
collisions), the heating (through the grain photoelectric mechanism), 
or the H$_2$ chemistry (through catalysis of H$_2$ on grain surfaces). 
Therefore, we have also considered a number of other models 
with different $r_{i,gas}$, but  we have not included dust 
since its effect is minimal.  
     
As we shall show in the next section, the line fluxes from the gas originate  
from the inner regions, $r_{i,gas} < r < r_{w,gas}$, where $r_{w,gas}$ 
is defined such that 
90\% of the line luminosity is generated from $r_{i,gas}$ to $r_{w,gas}$. 
Considerable (gas) opacity to the  stellar photons occurs at  
$r_{i,gas}$, so that $r_{w,gas}$ is typically only slightly ($\sim 10$\%) larger 
than $r_{i,gas}$,  leaving the shielded outer gas beyond $r_{w,gas}$  
cold and unemissive. Therefore, there can be very large amounts of cold gas  
in these outer regions, depending on the gas outer radius. 
The small amount of emission from this cold gas is below  
Spitzer line flux constraints. The IRS on Spitzer, however, is very  
sensitive to the warm gas just between  $r_{i,gas}$ and $r_{w,gas}$.  Because the emitting  
region has such small radial extent, we assume the gas surface density  
$\Sigma_{gas}(r) = \Sigma _0$ is a constant, independent of $r$ {\it in the emitting  
region}. We then vary $\Sigma _0$  
and compute the line fluxes.   
The observed line flux limits are then used to constrain $\Sigma _0$  
and the gas mass $M_w$ between $r_{i,gas}$ and $r_{w,gas}$.  
 
If the gas extends far beyond $r_{w,gas}$, the total mass of the gas in the disk  
can be estimated from   
\begin{equation}  
M_{gas}= {{2\pi\Sigma_0 r_{i,gas}^2}\over {2-\alpha}}\left[\left({r_{o,gas} 
\over r_{i,gas}}\right)^{2-\alpha}   
-1\right] \ \ {\rm for} \  \alpha < 2,  
\end{equation}  
where the gas surface density is given by $\Sigma_{gas}(r) = \Sigma _0(r_{i,gas}/r)^{\alpha}$  
for $r_{i,gas}<r<r_{o,gas}$.  Generally, $\alpha$  is assumed to be $\sim 0-1.5$. Note  
that for $\alpha <2$ and $r_{o,gas} >>r_{i,gas}$, considerable cold (hidden)  
gas mass is located at the outer radius $r_{o,gas}$.  
  
The gas emission line fluxes then depend on only two main parameters, $r_{i,gas}$ 
and $\Sigma _0$, in the case of HD~105 where dust does not affect the gas properties.  
The line fluxes vary with these parameters for  the following reasons.     
X-rays and UV photons tend to penetrate and heat a fixed column of gas.   
Therefore, larger inner holes increase the mass of  
gas relative to the total disk mass that is   
 affected by  X-ray and UV photons.  
However, at the same time,  the stellar radiation flux falls as $1/r^2$   
and the flux incident on the inner edge of the disk decreases as $r_{i,gas}$ increases.  
The total energy intercepted by the disk does not vary appreciably with changing 
$r_{i,gas}$, because the scale height tends to scale as $r$ so that the gas 
disk subtends a fairly constant solid angle.  However, the declining 
flux means a drop in the gas temperature (for fixed density), and the 
relative strengths of emission lines change.  Similarly, $\Sigma _0$ controls 
the gas density at $r_{i,gas}$, and increasing $\Sigma _0$ changes the gas 
temperature and the relative strengths of the lines.  We note that 
 at small enough $r_{i,gas}$ the lines will become undetectable 
by Spitzer regardless of $\Sigma _0$.  This arises because the mass and surface 
area of the emitting gas decline with $r_{i,gas}$.  Optically thin LTE line 
luminosities are proportional to mass, while optically thick lines are 
proportional to the emitting area.  At very small $r_{i,gas}$ and when 
 $\Sigma _0$ is raised to extremely high 
values, the lines cannot exceed their (optically thick) blackbody limits, the 
temperature is finite, and the small surface area and mass drive the beam diluted 
fluxes below detectability. 
 
We consider in the next section a range of $r_{i,gas}$ from 0.5 to 100 AU. We 
have chosen cases of 13, 19 and 45 AU because they correspond to the minimum, 
the best fit, and an approximation of the maximum $r_{i,dust}$ allowable 
from the SED modeling. 
However, we also consider the possibility that the gas does not co-exist with the 
dust.  In each case, we vary $\Sigma _0$ and determine the variation in 
the line fluxes with $\Sigma _0$ at the specified inner radius.

\section{Results and Discussion} 
\subsection{Gas Disks with Inner Holes} 
\begin{figure} 
\includegraphics[scale=0.6]{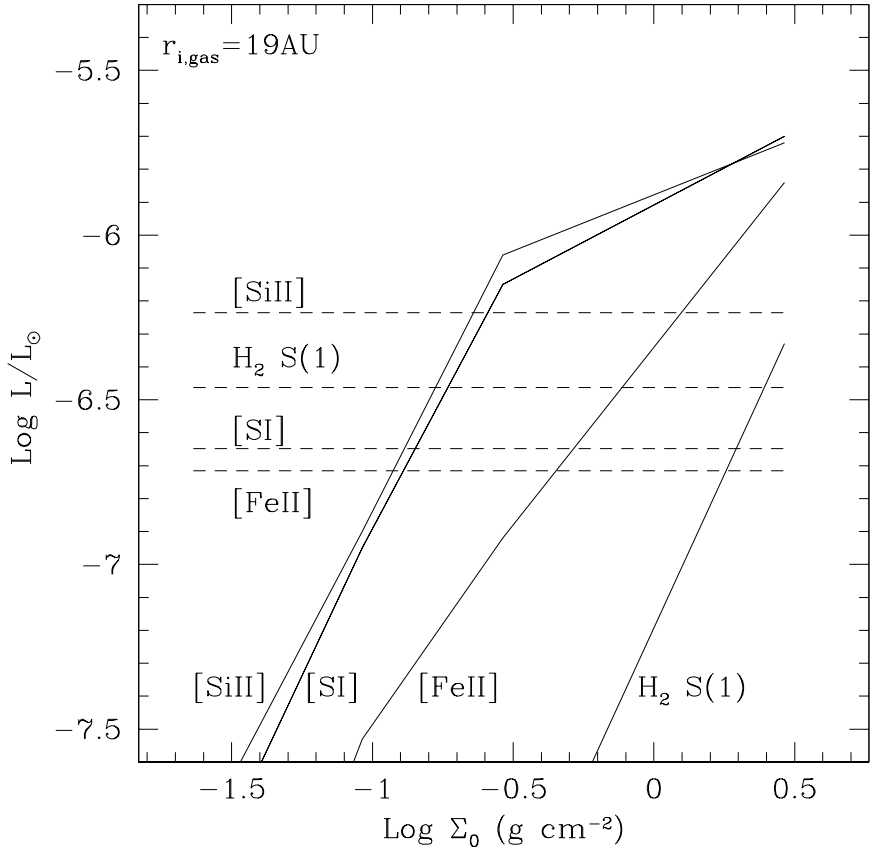} 
\caption{Calculated line luminosities (solid lines) 
as a function of gas surface density 
at the inner radius ($r_{i,gas}=r_{i,dust}=19$ AU) for the strongest 
lines obtained for the standard model disk with the best fitting 
dust to the SED. The dashed 
lines show observed upper limits on the  line luminosities 
as presented in this paper.  Where the dashed lines meet 
the solid lines for a given transition marks the upper limit 
to the gas surface density $\Sigma _{0,crit}$ in the disk 
as constrained by that transition.  In this case, [SI] sets 
the tightest constraint of $\Sigma _{0,crit} = 0.14$ gm cm$^{-2}$.} 
\label{gas} 
\end{figure}

We first discuss the results of the standard case where we have included 
the best fit dust population and $r_{i,gas}$ = $r_{i,dust}$ = 19 AU.     
Figure~\ref{gas} shows that model line luminosities   
initially increase as the gas mass surface density increases at the inner  
radius.  
The flux from the [SiII] 
line is the strongest followed by the [SI], [FeII],  
and the \mh S(1) line.  The [SiII] and [SI] lines are seen to plateau
at high surface densities as they reach their optically thick, blackbody
limits and as Si$^+$ begins to recombine at high density.
The dotted lines give the corresponding 
Spitzer upper limits presented in Table 1.  The intersection 
of the (dotted line) upper limits with the (solid line) model curves 
for a given species marks the critical column density $\Sigma _{0,crit}$ 
above which the line should have been detected.  With this particular 
choice of $r_{i,gas}$, the line providing the strongest constraint is the  
[SI] line, which gives $\Sigma _{0,crit} \simeq 0.14$ gm cm$^{-2}$.

It is instructive to look at the details of a fiducial model (the standard 
case at the critical surface density for that case, $\Sigma _{0,crit} = 0.14$ 
gm cm$^{-2}$) to understand the typical  
chemistry, temperature structure, and heating and cooling 
agents.    
\begin{figure} 
\includegraphics[scale=0.8]{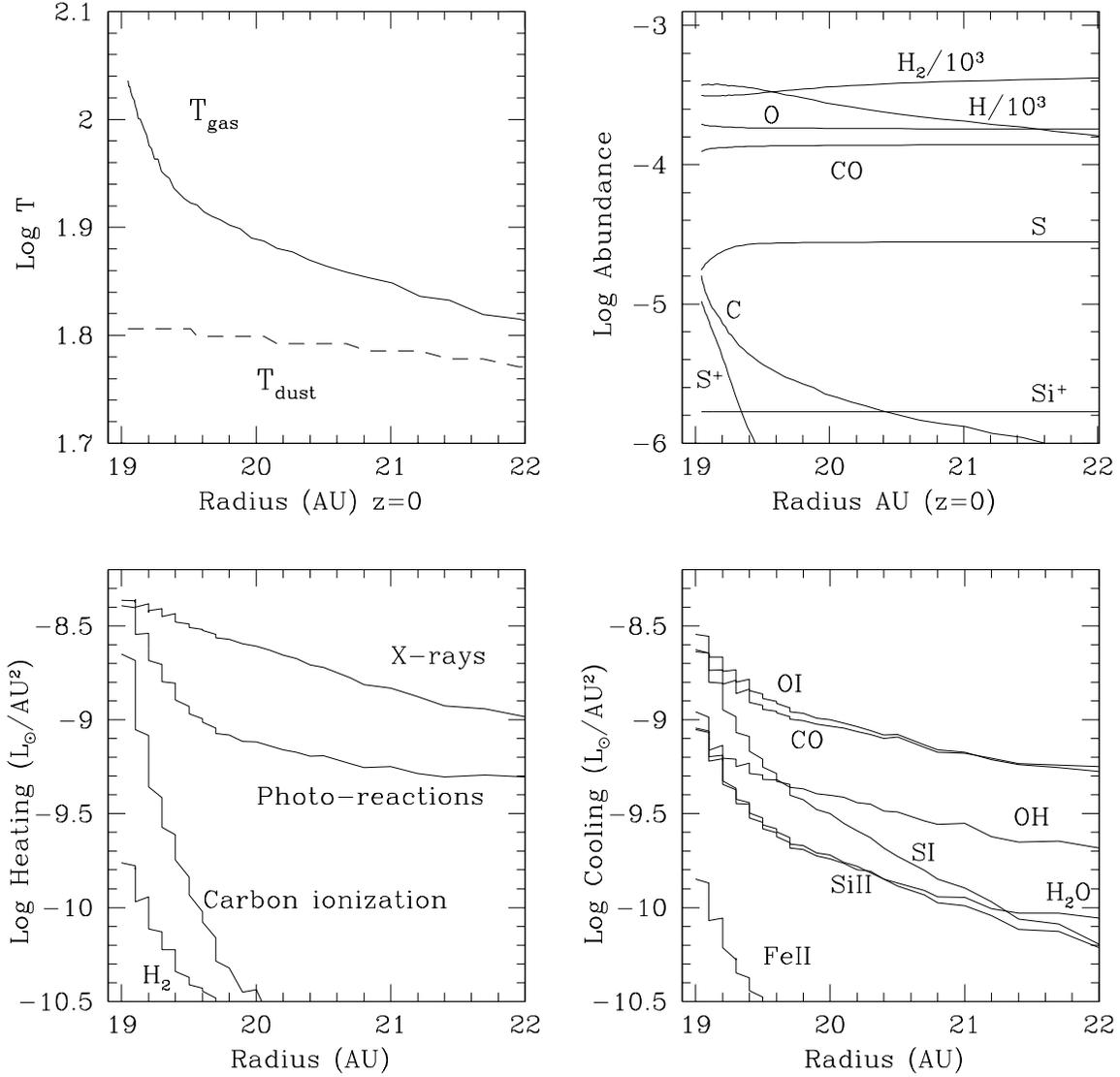} 
\caption{ The midplane gas and $a_{min}$ dust temperature, the midplane abundances of 
the abundant coolants,  the total vertically integrated disk heating, 
and total vertically integrated disk cooling for each species 
as a function of radius for the emission region extending from 
$r_{i,gas}$=19 AU to somewhat beyond $r_{w,gas}$. We have suppressed the 
data from 19 AU to 19.04 AU for clarity, as discussed in the text.}  
\label{disk-A} 
\end{figure} 
The upper left panel of Figure~\ref{disk-A} shows the gas temperature 
and the temperature of the smallest ($a_{min}=21$ $\mu$m) dust 
grains at the midplane as a function of $r$. 
The gas density at $r_{i,gas}$ in the midplane 
is about $10^9$ cm$^{-3}$. We have suppressed the data 
from 19 AU to 19.04 AU (a region where the gas goes from 
predominantly atomic hydrogen to mostly molecular hydrogen and where
where soft, $\sim 0.5$ keV X rays are absorbed), because of complicated behavior there which 
cannot be discerned at this graphic resolution. This small region does not 
significantly contribute to the line spectrum, since most of the emission arises from 
the much greater mass of gas that lies between 19.04 AU and 22 AU. 
At the inner disk edge (19 AU, not shown), the gas is atomic and fairly warm ($\sim 500$ K), 
but it very quickly cools to about 150 K at 19.04 AU because of the 
rapid rise in abundance of cooling molecular species such as CO. 
  One sees that beyond 19.04 AU 
the temperature drops from $\sim 150$ K to about 70 K at 22 AU. 
The chemistry changes slowly in this region.  The hydrogen  
is mostly H$_2$, formed by 
the reaction of H with H$^-$ and on grain surfaces. We have 
found that eliminating grains does not change the H$_2$ abundance 
appreciably (it is nearly entirely molecular anyway). The atomic 
H abundance is quite high because of the inefficiency of the H$^-$ 
and grain processes.  Most of the sulfur is atomic, most of the silicon 
and iron is  singly ionized.  The carbon is mostly in CO, and all 
remaining gas phase oxygen is in atomic O. 
 Note that when $r=$ 19.1 AU, there is already 
a column $N\sim 10^{22}$ cm$^{-2}$ of hydrogen between the central star 
and that point.  Because the densities 
are very high, and the UV field low, there is considerable H$_2$,  
C, CO, S, Si, and Fe 
in the surface regions, and their column provides considerable opacity 
to the stellar UV photons via photoionization and photodissociation processes. 
The dust surface area per hydrogen atom  
is small, so that dust extinction is not important, and the opacity is 
entirely due to gaseous species.  Similarly,  the grain surface 
area is too small to effectively heat the gas by the 
grain photoelectric heating mechanism, or to heat or cool the 
gas by gas-grain collisions.  The gas-grain cooling could not be 
shown in the lower right panel because it is many orders of magnitude 
below the graph. We have also run alternate SED-fitting models 
of allowable $r_{i,dust}$ and found that  dust is not 
important in HD~105 for modeling the gas in the disk because of its low 
abundance and surface area. 
  
The main heating mechanisms for the gas in the emission zone are the 
heating by gaseous absorption of X-rays from the central star and of  
the stellar  
optical and UV 
photons which photoionize and  
photodissociate atoms, ions and molecules.   
 The lower energy X-ray photons are absorbed  
nearest 
to the surface and, because of their higher cross sections for   
absorption,  
lead to the highest X-ray heating there.  The more energetic X-rays  
penetrate 
further, and provide gas heating at greater depths. 
In the bulk 
of the emissive zone, the heating by optical and UV photons 
is dominated by the photoionization 
of S (which provides most of the electrons) and the photodetachment 
of electrons from H$^-$. Note that this latter process requires only  
stellar optical photons and not the higher energy UV photons.  However, 
this process requires H$^-$; the H$^-$ is produced by electrons 
which arise from the photoionization of S and Si; and these 
photoionizations do require UV photons.  
The cooling is mainly by CO mid J rotational 
transitions ($\sim 200 - 400$ $\mu$m), [SI] 25 $\mu$m, 
and [OI] 63 $\mu$m.  Beyond 22 AU, the cooling is dominated by CO mid J  
transitions, 
as the gas cools below 70 K. If the gas disk extends much beyond 
30 AU, these CO transitions become detectable if the CO does not 
freeze onto the cold grain surfaces. To date, no CO observations 
of HD~105 have been made.  We note that the cooling by CO and  [OI] 
occurs at longer wavelengths ($\lambda > 38$ $\mu$m) than accessible 
by the IRS on Spitzer, but that these transitions may be detectable 
by instruments on the future Stratospheric Observatory for Infrared 
Astronomy (SOFIA) and by the Herschel Observatory.   

Direct heating of the gas by absorption of stellar photons 
heats the gas to $T\gtrsim 70$K in the inner regions (19-22 AU) of
 the gas disk. The gas 
vertical scale height is proportional to $T^{1/2}$, and the warm gas  
intercepts 
about 20\% of the stellar radiation due to its flared nature.  
In many wavelength bands the gas opacity is significant. Therefore, 
there is significant heating by this mechanism, but it tends to 
occur at the inner rim ($r_{i,gas}$), where the stellar fluxes are highest and least 
attenuated by the gas. The heating at the inner rim makes it expand 
vertically, enhancing its ability to intercept photons and to shield 
the outer disk.  Dullemond, Dominik \& Natta (2001) have discussed this effect for 
dusty disks where the opacity is provided by the dust particles. 
We find  in the fiducial case that 
heated region extends radially about 3  AU beyond $r_{i,gas}$ to  
$r_{w,gas}\simeq 22$ AU.  
Of order 90\% of the luminosity of the infrared lines studied here is emitted 
in this narrow annulus. 
Because of this fact, the gas spectra from the model disks around HD~105, where 
dust plays no factor, depend only on $r_{i,gas}$ and the gas surface density 
$\Sigma _0$ there, as discussed above. 
 
We now present the results of an identical search of $\Sigma _0$ parameter space 
for a variety of $r_{i.gas}$ from 0.5 to 100 AU, assuming dust is unimportant. 
Table~\ref{masses} lists the surface density limits $\Sigma _{0,crit}$ as determined from each   
line.  Associated with each  
of these lines and $\Sigma _{0,crit}$ are  particular values for $r_w$  
and $M_w$, 
the mass of the warm gas between $r_{i,gas}$ and $r_w$.  These are also listed in 
Table~\ref{masses} for the most sensitive line.  Note that the warm gas masses  
are somewhat less than the simple  
analytic expression 
given in \S 4.1 for H$_2$ lines produced in $T \sim 100$ K gas.  Spitzer can 
detect masses somewhat smaller than $10^{-2}$ M$_J$ in HD~105 (the limit derived 
from H$_2$ S(1) 
for $T\sim 100$ K gas) because other lines such as [SI] are predicted to be stronger, 
and, in the cases of small $r_{i,gas}$, the gas is warmer than 100 K. 
 
\begin{table}[t] 
\caption{$M_w$, $r_{w,gas}$,  and $\Sigma _{0,crit}$  
as derived from the observations and models} 
 
\begin{tabular}{lrrrrrrr} 
\hline 
$r_{i,gas}$ (AU) & 0.5   &  1.0  & 5.0   & 13.0  &  19.0  & 45.0  & 100.0 \\ 
\hline 
$r_w$ (AU)       & 0.69  & 1.28 &  5.15 & 16.5 & 21.9  & 48.6 & 123.0    \\ 
                 &       &      &       &      &       &      &     \\ 
$\Sigma_{0,crit}$ (g cm$^{-2})$ H$_2$ S(1)  
                 & --    &  --  & 251.10 & 2.30  & 1.82  & 6.39 &  2.98   \\ 
$\Sigma_{0,crit}$ (g cm$^{-2})$ [FeII] 
                 & 6.12  & 2.32 & 0.92 & 0.45 & 0.45  & 0.42 & 1.31    \\ 
$\Sigma_{0,crit}$ (g cm$^{-2})$ [SI] 
                 & 3.43  & 0.36 & 0.12 & 0.11 & 0.14  & 0.21 &  1.04   \\ 
$\Sigma_{0,crit}$ (g cm$^{-2})$ [SiII] 
                 & 121.30 & 21.4 & 9.35 & 0.26 & 0.23  & 0.41 & 2.02    \\ 
                 &       &      &      &      &       &     &     \\ 
Log $M_w$ (M$_J$)& -3.57 & -4.09& -4.19 & -2.41 & -2.22 & -1.61 & 0.6 \\ 
\hline 
\end{tabular} 
\label{masses} 
\end{table}

The total mass of gas (warm and cold) in any given model depends on the power law of the gas 
surface density distribution $\alpha$ and on the outermost extent $r_{o,gas}$ 
of the gas.  Let us assume that the region of gas giant formation extends 
to about 40 AU, based on the solar system example.  
We take one of the more extreme values of the power law of the gas surface density  
distribution,  $\alpha =0$, which maximizes the gas mass in the outer shielded zones. 
The upper limit (derived from the most sensitive line [SI]) to the total 
gas mass in this extreme case of low $\alpha$ 
is then about 1.9 M$_J$ for r$_{i,gas}=0.5$ AU,  0.2 M$_J$ for r$_{i,gas}=1$ AU, and 
0.1 M$_J$ for r$_{i,gas}=5-20$ AU,  
where we basically extrapolate the upper limit 
on the gas surface density measured at $r_{i,gas}$ to 40 AU.  
  {\it Therefore, for $r_{i,gas} \gtrsim 0.5$ AU, there is insufficient 
gas in HD~105 at this time to feed the formation  of gas giants}.

Table 2 shows that the upper limits to the gas surface density at $r_{i,gas}$ 
are not very 
sensitive to $r_{i,gas}$ for 1 AU $\lesssim r_{i,gas} \lesssim 40$ AU.  Larger $r_{i,gas}$  
tends to produce more mass of gas heated by 
the stellar photons, but, because of the dilution of the stellar flux, also  
tends 
to lead to lower characteristic gas temperature. These effects counterbalance  
each 
other to some extent, resulting in similar line fluxes.  
However, for $r_{i,gas} < 1$ AU, $\Sigma _{0,crit}$ rises steeply.  
The heating mechanisms only penetrate to a relatively constant column of $N_w \sim  
10^{22}$ cm$^{-2}$ 
so the mass of heated gas scales roughly as $r_{i,gas}^2$, assuming the vertical  
scale height 
scales with radius.  Therefore, the mass (or area for optically thick lines) 
of heated gas goes down as we move inward, 
and so the gas temperature must rise appreciably in order for the gas to be  
detected. 
Since the gas temperature tends to rise with increasing density, due to the  
collisional 
de-excitation of the upper levels of cooling transitions, we require  
substantially  more 
surface density to raise the temperature sufficiently to overcome the loss of  
warm gas mass and area. 
%
  
\begin{figure} 
\includegraphics[scale=0.5]{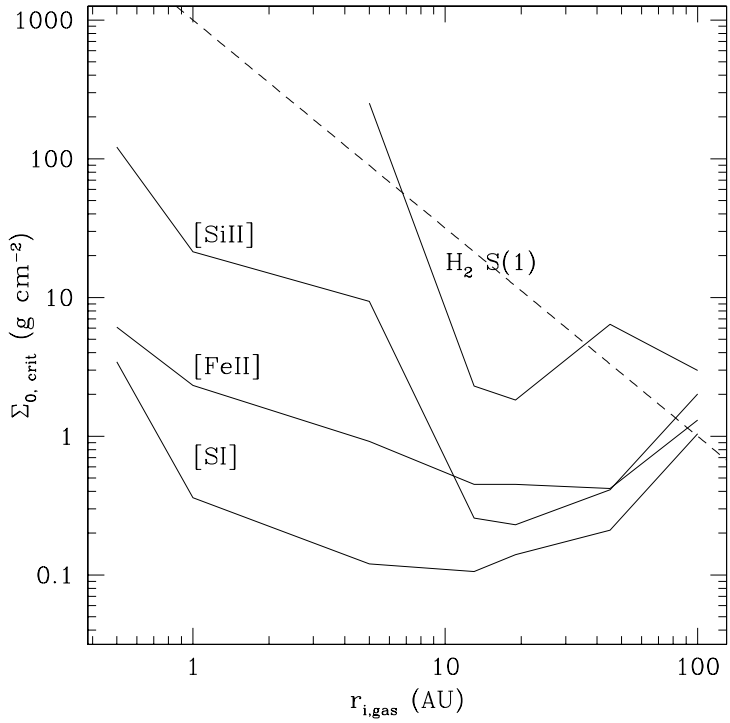} 
\caption{ The detectable surface density based on Spitzer observations of HD~105 
for a gas disk with various inner radii, and with the small amount of dust 
implied by the infrared continuum SED.  For small $r_{i,gas} \lesssim 1$ AU, the lines 
set poor or no limits, as the warm emissive gas gets too optically thick and beam 
diluted to be detectable by Spitzer, regardless of the magnitude of $\Sigma _0$ (see text). Also shown for comparison is the surface density distribution of the minimum mass solar nebula with radius,  
$\Sigma(r)=1000 r_{AU}^{-3/2}$ g cm$^{-2}$ (dashed line).}  
\label{sigma-rin} 
\end{figure}

Figure~\ref{sigma-rin} visually shows how the critical surface density  
$\Sigma _{0,crit}$ depends on  
$r_{i,gas}$.  We see that [SI] 25 $\mu$m is the most sensitive 
Spitzer line for constraining the gas surface density in gas disks 
with little dust, like HD~105. 
If $r_{i,gas}$ is sufficiently small ($\lesssim 0.5$ AU), the gas mass in the  
disk 
is essentially unconstrained by the Spitzer observations because a very small  
mass 
and surface area of inner gas shields the outer regions, leaving them too cold 
to be detected at this time.  Clearly, if a line detection is made 
by Spitzer IRS, knowledge of the parameter $r_{i,gas}$ is extremely 
helpful if the line flux is to be 
converted into $\Sigma _{0,crit}$ and $M_w$. Therefore, followup observations 
by higher spectral resolution instruments will be very useful, since resolved line widths 
can be translated into Keplerian velocities and, hence, $r_{i,gas}$. 
 
\subsection{Discussion} 
 
Assuming that $r_{i,gas} \gtrsim 0.5$ AU,  
the Spitzer upper limits 
to the gas lines fluxes in [SI] 25 $\mu$m, [SiII] 35 $\mu$m, [FeII] 26 $\mu$m,  
and 
H$_2$ S(1) 17 $\mu$m provide upper limits to the gas surface density at $r_{i,gas}$ 
which would indicate less than a Jupiter mass of gas is present in the 
gas giant planet-forming zone which extends to 40 AU. It is interesting 
that the most sensitive indicators of gas are not the H$_2$ lines, but fine  
structure 
lines (in particular [SI] 25 $\mu$m) of less abundant species  
(see Gorti \& Hollenbach 2004). These upper  
limits indicate that 
in this region of possible gas giant planet formation, either a giant planet
has already formed or it never will: {\it the era of significant  
gas accretion onto protoplanets  
is over for this 30 Myr old system}.   
 
The main caveat is that if the gas surface  
density distribution, 
unlike the dust, extends to the innermost regions ($r_{i,gas}<  
0.5$ AU) of the disk, 
then significant (i.e., $> 1$ M$_J$) amounts of cold gas could be hidden in 
the 5-40 AU region where gas giant planets may still be accreting, without violating 
our observational constraints on the line luminosities.  
However, if the gas  indeed extends 
to $< 0.5$ AU, then the gas is likely accreting onto the stellar surface. 
 If $M_d$ is the mass of gas from the stellar surface to  40 AU, then the accretion rate onto 
the central star is given $\dot M \simeq 3\times 10^{-9}(\alpha/0.01)(M_d/$M$_J$) M$_\odot$ 
yr$^{-1}$, where $\alpha$ is the standard alpha viscosity 
parameter.  Such high accretion rates would  
 produce diagnostics, such as large UV or near-IR excesses, H$\alpha$ emission lines, 
or other indicators of accretion-driven winds, which are not observed.  Therefore, 
we regard this possibility as unlikely. 
 
The gas limits in the 10-40 AU region may also be relevant to theories of  
the formation 
of the outer  giants such as Neptune and Uranus.   
A generic problem for  
these 
systems is that they heat dynamically (i.e., induce high random velocities) the much smaller  
objects 
which provide potential coalescent collisions that enable them to grow.  The  
higher 
velocities reduce the gravitational focusing, and therefore reduce their growth  
rate (Levison 
and Stewart 2001).  Without some kind of dynamical cooling mechanisms, it is  
difficult for 
them to form in situ within the lifetime of the solar system.  Gas drag is one  
possibility, but it requires many Jupiter masses of gas.
Our upper limits on the amount of gas indicate that, after 30 Myr in HD~105,  
there is  not nearly 
enough gas drag to have an appreciable effect (see, e.g., Goldreich et al. 2004)
at this time. 
 
Takeuchi and Artymowicz (2001), Klahr \& Lin (2001), and Takeuchi \& Lin (2002) 
 showed that a small amount of  
gas (and 
the effect of stellar photons on the gas and dust) can 
sculpt the dust morphology and create an inner hole. Therefore, the sharp inner  
dust hole at $\sim 19$ AU in HD~105  need 
not reflect the presence of giant planets, but may be produced in a planet-less 
gas/dust disk around HD~105. 
These authors show that $M_{gas} > 1-10 M_{dust}$ produces such effects, and that when  
$M_{gas} > 
100 M_{dust}$ the dust grains cannot migrate effectively relative to the gas.  
Using our 
upper limits for the case where the gas and dust are co-spatial, we find that we  
can only 
constrain the gas mass to dust mass ratio in HD~105  
to values of $< 1000$, and that therefore we  
cannot 
effectively constrain the effects of gas on the dust dynamics.  In other words, the  
sharp 
inner rim of dust implied by the IR continuum SED of HD~105  
could be produced by a small  
amount 
of gas below our upper limits. An alternate hypothesis is that in the
absence of gas, a giant planet is preventing dust from the outer debris
disk from reaching the inner part of the system, as discussed in M04.
 
Kominami \& Ida (2002, 2004) and Agnor \& Ward (2002) discuss the effects of a  
small amount 
of gas in the terrestrial zone (1-5 AU) on the resulting formation of terrestrial  
planets.  If 
$M_{gas} >> 10^{-2}$ M$_J$ in the terrestrial zone 
for tens of millions of years, then lunar and Mars-sized 
planetary embryos feel the dynamical friction of the gas, circularize their  
orbits, 
and never collide to form Earth-mass planets.  On the other hand, if $M_{gas} <<  
10^{-2}$ 
$M_J$, then the embryos are on eccentric orbits, and collide to form  
Earth-sized or 
larger orbits, but with eccentricities substantially larger than the Earth.  The  
suggestion 
is that the Earth may have formed with roughly 10$^{-2}$ M$_J$ of gas in the  
terrestrial zone 
for tens of millions of years.  The terrestrial zone ($\sim 0.3-3$ AU) in HD~105 is  
inside 
$r_{i,dust}$ and is currently quite dust-free.  Our models  
show that if the gas extends all the way  in to  
$r_{i,gas} < 0.5$ AU, then we are not sensitive to gas mass in the terrestrial  
zone, and therefore 
cannot set useful limits on this process.  If for some reason the gas had  
an inner radius 
of 0.5 -- 1 AU, we can set limits of about $2\times 10^{-2}$ to $2\times 10^{-3}$ M$_J$ for the 
mass of gas from this inner radius to 5 AU. 
In this case, the limits are close to the critical value of $10^{-2}$ M$_J$, and 
indicate perhaps insufficient gas to prevent large, Earth-sized planets from 
forming.   
Spitzer may be able to set even more stringent limits on this  
process in other 
sources.  If sufficient small dust is mixed with the gas in the terrestrial  
zone, then the 
gas in the models is hotter, and smaller gas masses can be detected. 
 
Finally, are these upper limits surprising in the context of theoretical  
calculations 
that have been performed on the likely dispersal mechanisms of the gas?   
The current kinematic evidence suggests that HD~105 formed in a small group 
of tens of stars (see \S 2), which likely lacked O or early B-type stars. 
Assuming that HD~105 was not exposed to high UV fluxes from nearby massive stars, 
the main dispersal mechanism for the outer disk would likely be photoevaporation  
of the gas by the central 
star, and for the inner disk 
viscous accretion and spreading of the gas  would dominate (Hollenbach, Yorke 
\& Johnstone 2000, Clarke et al. 2001, Matsuyama  et al. 2003).  Unfortunately, 
it is difficult to answer this question because of the lack of self consistent 
photoevaporation models which treat not just the EUV (i.e., $h\nu > 13.6$ eV) 
photons (see Hollenbach et al. 1994), but also the less energetic  
stellar photons as well as the X rays  
from the young central star. 
Gorti \& Hollenbach (in preparation) are developing such models, and their 
preliminary results show 
that the less energetic photons rapidly (in less than 10 Myr) disperse the gas 
outside of about 30-50 AU.  The EUV photons can potentially remove gas outside 
of about 1-5 AU, but the evolution of the EUV luminosity and the radiative 
transfer of these photons as they try to penetrate the protostellar winds 
in the early stages of star formation are not well determined.  A recent 
paper by Alexander, Clarke \& Pringle (2005) suggests rather high  
escaping EUV luminosities which may 
rapidly ($< 10$ Myr) remove the outer gas.  Once the outer disk is
removed, viscosity removes the inner gas 
on timescales which are roughly 0.1 Myr (0.01/$\alpha _v)(r/$10  
AU), where $\alpha _v$ is the turbulent viscosity parameter in the  
standard ``$\alpha$'' disks. 
The 
value of $\alpha _v$ 
is typically 10$^{-2}$ if the Balbus \& Hawley (1991) magneto-rotational 
instability, or MRI, is operant.  The MRI instability requires a minimal level 
of ionization, which all our relatively low mass models meet.  
  Assuming  that MRI is active, the viscous 
timescales inside the photoevaporation region (which lies at $>3 - 30$ AU) 
are of order 0.03 -- 0.3 Myr --extremely short! On the other hand, it is not 
totally certain that this instability would be fully active (Chiang, Fischer, \& Thommes 
2002).  
Therefore, 
we conclude that these upper limits are not surprising, but they do set  
constraints 
on the rather poorly known dispersal mechanisms.

\section{Summary and Conclusions} 
 
      One of the goals of the Formation and Evolution of Planetary Systems 
 (FEPS) Spitzer Legacy project is 
to measure the evolution and dispersal of gas 
in the planet-forming regions ($\sim 0.5 - 40 $ AU) of disks 
around solar-type stars of ages 3--100 Myr. We report here 
our first carefully reduced and analyzed high spectral resolution data 
taken by the IRS instrument on the Spitzer Space Telescope. This paper 
illustrates our method of modeling the data to 
obtain constraints on the gas surface density distribution and mass. 
We eventually plan to obtain data on about 40 nearby stars in this 
age range to look for variation with age and other stellar properties. 
 
     The data presented here are for the source HD~105, a $\sim 30$ Myr 
old G0 star at a distance of 40 pc with a known IR excess arising from 
a circumstellar dust disk orbiting at $r_{i,dust} \gtrsim 13$ AU. The 
derived upper limits to the H$_2$ S(0) 28 $\mu$m, H$_2$ S(1) 17 $\mu$m, 
H$_2$ S(2) 12 $\mu$m, [SI] 25 $\mu$m, [FeII] 26 $\mu$m, and [SiII] 35 $\mu$m 
lines are given in Table 1.  The H$_2$ upper limits directly place limits 
on the mass of warm gas in the disk: $M({\rm H_2}) \lesssim 4.6$ M$_J$ 
at 50 K, $3.8\times 10^{-2}$ M$_J$ at 100 K, and 3.0$\times 10^{-3}$  
M$_J$ at 200 K.   This can be compared with the roughly 10$^{-3}$ M$_J$ 
of gas detected around $\beta$ Pictoris,  
a 10-20 Myr old A5 star at a  
distance of about 19 pc (Brandeker et al 2004). It can also be compared
with the recent UV absorption measurements and analysis of AU Mic, an
M1 star in the $\beta$ Pictoris Moving Group at a distance of 9.9 pc
and with a similar age as $\beta$ Pic (Roberge et al 2005).   They
find an upper limit to the H$_2$ mass of about $2\times 10^{-4}$ M$_J$.
It appears that even in the inner ($< 30$ AU) regions not well
probed by CO observations, the gas is largely dissipated in
these three sources which span the age range 10-30 Myr, and which
span the stellar types from M1 to A5.

  Detailed thermal/chemical models of HD~105 were constructed and 
compared with the Spitzer observations to obtain further  
constraints on the gas mass and surface density.  These models 
calculate the gas temperature, chemistry, and vertical structure 
self-consistently and predict line fluxes which depend largely on 
the gas inner radius, $r_{i,gas}$, and the gas surface density 
$\Sigma _0$ there.  We show that most of the gas emission arises 
in a thin inner rim, extending from $r_{i,gas}$ to $r_{w,gas}$, heated 
by stellar optical, ultraviolet, and X-ray photons. 
The upper limits 
on the [SI] 25 $\mu$m and [SiII] 35 $\mu$m fine structure lines provide 
the strongest constraints on 
the gas surface density $\Sigma _0$ at the inner rim. 
 
If the gas inner radius is comparable to the observationally constrained 
dust inner radius, $r_{i,dust} \gtrsim 13$ AU, we show that the Spitzer 
upper limits on the line fluxes limit $\Sigma _0$ to $\lesssim 0.2$ 
gm cm$^{-2}$ for $r_{i,gas}$ between 10 and 40 AU. In this case, the total 
mass of gas (cold and warm)  
in the planet forming region between 10 and 40 AU is constrained to be 
less than 0.1 M$_J$ (assuming a constant surface density between 
$r_{i,gas}$ and 40 AU). 
 
The gas may not co-exist with the dust, however.  We show that even if 
the putative gas extends inward to $r_{i,gas}=0.5$ AU, the Spitzer upper 
limits set tight constraints on the gas surface density and mass. 
The upper limits on the gas surface density are $\Sigma _{0,crit}$ =  
3.43, 
0.36, and 0.12 gm cm$^{-2}$ 
for $r_{i,gas}$ = 0.5, 1, and 5 AU, respectively.  The total mass 
of gas in the gas giant planet-forming region out to 40 AU depends on 
how we extrapolate the gas surface density from the inner radius. 
If we assume a gas surface density power law $\Sigma \propto r^{-3/2}$, 
which is often assumed in disks, then the total gas mass in these 
three cases is limited to about $10^{-2}$, 3$\times 10^{-3}$, and  
$10^{-2}$ M$_J$, respectively.  An 
extreme assumption, allowing for the most hidden cold gas mass,  
would be a constant surface density with radius.  In this case, 
we obtain total (warm and cold) 
gas mass limits of 1.9, 0.2, and 0.07 M$_J$, respectively. 
In summary, assuming that $r_{i,gas} > 0.5$ AU and any reasonable 
gas surface density distribution, there is less than a Jupiter mass 
of total gas in the gas giant planet forming region out to 40 AU. 
Given likely temperature 
distributions produced in our models, Spitzer is unlikely to be able 
to detect {\it total} gas masses less than about 10$^{-2}$ M$_J$ 
for disks around relatively nearby ($\sim 30 pc$) low mass stars.

If the gas extends to $r_{i,gas} \lesssim 0.5$ AU, the Spitzer upper 
limits set no useful constraints on the gas mass, because gas lines 
become undetectable even with extremely large gas masses.  In this 
case, the small inner rim which absorbs the heating photons from the star 
has too little mass (for optically thin lines) and too little surface 
area (for optically thick lines) to provide detectable Spitzer emission, 
regardless of the magnitude of $\Sigma _0$. The outer gas is effectively 
shielded from the heating photons, and significant cold gas mass could 
exist in the planet forming regions.  We argue that this case is 
unlikely, however, since the gas would likely extend all the way to 
the stellar surface, and viscous accretion onto the central star 
would lead to observational diagnostics such as near infrared or UV 
excess, H$\alpha$ emission, and signs of winds generated by the 
accretion process (which are not observed). 
 
We therefore conclude that the Spitzer upper limits imply low 
upper limits to the gas surface density and mass in the 0.5-40 AU 
region around HD~105.  We also note that, like HD~105, 
 many debris disks will have too little dust to affect the gas spectra;
therefore, a similar analysis of Spitzer upper limits on line fluxes
will result in similar conclusions. As discussed in \S 5, the limits are not  
sufficiently stringent to constrain 
the potential effects of gas on dust dynamics.  They set interesting 
limits on the mass of gas which might affect terrestrial planet formation 
only in the {\it ad hoc} case where the inner gas radius is of order 
0.5 -- 1 AU. However, the limits do set interesting constraints for 
giant planet formation.  The upper limits to the gas mass obtained here 
are too small 
to enhance the buildup of gas poor outer giants such as Uranus or Neptune, 
and too small to allow for gas giants to form from 
the gas reservoir after this time ($\sim 30 $ Myr).    
 
We acknowledge support from NASA's Spitzer Space Telescope Legacy program, 
which has supported our group, the Formation and Evolution of Planetary 
Systems Legacy team. We thank the rest of the FEPS team for 
their efforts in making the FEPS project successful, and for their
help in obtaining, analyzing, and interpreting the data discussed
in this paper.

\end{document}